\documentclass[10pt,reqno]{amsart}

\pdfoutput=1

\usepackage{graphicx}

\begin{document}

\title{On probabilities of Risk type board game combats}

\author{Manu Harju}
\email[]{manu.harju@tut.fi}
%\homepage[]{http://www.cs.tut.fi/~harju23/}
%\thanks{}
%\affiliation{}

%\date{\today}

\begin{abstract}
Risk is a well-known turn based board game where the primary objective is nothing less than the world domination. Gameplay is based on battles between armies located in adjacent territories on the map of Earth. The combat's outcome is decided by rolling dice, and therefore a probabilistic approach can be taken. Although several results are derived, the conclusions suggest that the gameplay is highly depending on luck.
\end{abstract}

%\keywords{dice, probability, risk, board game, game theory}

\maketitle

\section{\label{sec:intro}Introduction}
Risk\cite{lamorisse:risk} is a classic strategic board game where the players aim to conquer the world. The original version of the game was released in 1957. The board consists of a map of the Earth and the gameplay is based on combats between players' troops in neighboring territories. Territories must be adjacent or connected by a sea line on the board. 

The studies of this paper concentrate on the revised version\cite{daviau:risknew} of the original game. In the combat the attacker has at least one and at most three attacking troops attacking from one territory to another. Nevertheless, one can attack from the same territory to a different territory if there is enough troops, and from different territories to the same target territory. The combat rules are relatively simple. Players are throwing ordinary six sided dice, and the number of dice of both players is corresponding to the number of troops attacking and defending, though the attacker is throwing at most three and the defender at most two dice. By the rules the defending player throws dice after attacker's throw. However, this doesn't change the probabilistic analysis of the combat in any way. When the dice are thrown, the best dice are compared and the bigger one wins. In the case of tie, the defender wins. If the both players had at least two dice, then the second best are compared. Losing one die means losing one troop, so the biggest loss on one round of throwing can be only two troops \cite{daviau:risknew}. However, a player can continue the attack as long as he/she wants to. In this paper we assume that the attacker continues until the territory is conquered or the defender kills all attacking troops.

The study of this paper concentrates on the probabilities of conquering a territory. Several concepts are presented to ease the handling of several rounds of combat and to avoid too complicated combinatorial formulas. However, when calculating the distributions these can not be avoided. A few different situations are investigated, like the more interesting case of an attack from several different territories. In addition, the expectation of lost troops in a successful attack is studied, as well as the expectation of surviving defending troops.


\section{\label{sec:theory}Theory}
%\section{\label{sec:ordstat}Order statistics}
\subsection{\label{sec:ordstat}Order statistics}
Let $X_1, \ldots, X_n$ be $n$ independent and identically distributed random variables with cumulative distribution function $F(k)$ and probability mass function $f(k)$. The $k$th order statistics\cite{hogg:intro} i.e. the $k$th-smallest value of the sample $X_1, \ldots, X_n$ is denoted by $X_{(k)}.$

Now every $X_{(k)}$ is a random variable, and for every $k$ the probabilities can be written as
\begin{eqnarray}
\label{eq:p1}P(X_{(k)} \leq m) & = & \sum_{j=0}^{n-k} {n \choose j} (1 - F(m))^j F(m)^{n-j} \\
\label{eq:p2}P(X_{(k)} < m) & = &
\sum_{j=0}^{n-k} {n \choose j} (1 - F(m) + f(m))^j (F(m) - f(m))^{n-j} \\
\label{eq:p3}P(X_{(k)} = m) & = & P(X_{(k)} \leq m) - P(X_{(k)} < m).
\end{eqnarray}

Since we are playing with ordinary 6-sided dice we have 
\begin{eqnarray}
f(m) & = & \left\{ 
\begin{array}{rl}
	\frac{1}{6} & \text{if } 1 \leq m \leq 6 \\
	0 & \text{otherwise,}
\end{array}\right. \\
F(m) & = & \left\{
\begin{array}{rl}
	0 & \text{if } m < 1, \\
	\frac{m}{6} & \text{if } 1 \leq m \leq 6, \\
	1 & \text{if } m > 6.
\end{array}\right.
\end{eqnarray}

Substituting these to Eq. (\ref{eq:p2}) and (\ref{eq:p3}) we get
\begin{eqnarray}
P(X_{(k)} < m) & = & \sum_{j=0}^{n-k}{n \choose j} (\frac{7 - m}{6})^j (\frac{m-1}{6})^{n-j} \\
P(X_{(k)} = m) & = & \sum_{j=0}^{n-k}{n \choose j} \left[(1 - \frac{m}{6})^j (\frac{m}{6})^{n-j}\right. 
 - \left. (\frac{7-m}{6})^j (\frac{m-1}{6})^{n-j}\right].
\end{eqnarray}

In our case we are only interested in two of the order statistics, namely $X_{(n)} = max\{X_i\}$ and $X_{(n-1)}$. These can be written as
\begin{eqnarray}
\label{eq:ekap}P(X_{(n)} < m) & = & (\frac{m - 1}{6})^n, \\
\label{eq:p4}P(X_{(n)} = m) & = & % (\frac{m}{6})^n - (\frac{m-1}{6})^n \nonumber \\ & = &
\frac{m^n - (m-1)^n}{6^n}, \\
P(X_{(n-1)} < m) & = & \nonumber \\ 
(\frac{m - 1}{6})^n & + & n (\frac{7 - m}{6})(\frac{m - 1}{6})^{n - 1} \\
\label{eq:p5}P(X_{(n - 1)} = m) & = & (\frac{m}{6})^n - (\frac{m-1}{6})^n \nonumber \\
+ n [(1 - \frac{m}{6})(\frac{m}{6})^{n-1} & - & (\frac{7 - m}{6})(\frac{m-1}{6})^{n-1}].
% & = & \frac{1}{6^n}\left[m^n - (m-1)^n + n (6m^{n-1} - m^n - (7-m)(m-1)^{n-1} ) \right].
\end{eqnarray}

If the attacker has $n_a$ dice and the defender $n_d$ dice, the probability that the attacker wins the best dice can be expressed as
\begin{equation}
	% \sum_{m=2}^6 P(X_{(n)} = m) P(X_{(n)} < m) = 
p = \sum_{m=2}^6 \frac{m^{n_a}(m-1)^{n_d} - (m - 1)^{n_a + n_d}}{6^{n_a + n_d}}.
\end{equation}

%\section{\label{sec:comseq}Combat sequences}
\subsection{\label{sec:comseq}Combat sequences}
In a standard game we need Eq. (\ref{eq:ekap}) - (\ref{eq:p5}) with $n$ equal at most 3 (attack) or 2 (defense). If a player has obtained the attack (or defense) die, the maximum number of dice is just increased by one. However, still the best two dice only are compared.

Since the maximum number of units that can attack one territory once is three, there might be several rounds in the combat. Although in the game the attacker can cancel the attack at any point, we consider only the case that either the attacker or the defender loses all his/her troops. 

Now consider the case where the attacker has three troops and the defender has five troops. When calculating the total probability that the attacker wins, all possible sequences leading to the desired result must be taken into account. For example, ``\emph{on the first round the defender loses two troops, on the second round both lose one troop, and on the third round the defender loses two troops}'' can be considered as one combat sequence. All those sequences where the attacker wins can be drawn into a rooted tree, where the weight of an edge corresponds to the number of units the defender loses, and the path length from the root to any leaf is exactly the number of troops the defender had in the beginning. From the tree it should be reasonably easy to compute the probability of all sequences, and the total probability that the attacker wins is just the sum of those probabilities.

%Formally the probability that the attacker wins can be presented as follows. Let $X_1, \ldots, X_{n_1}$ be the random variables corresponding to the attacker's dice, and $Y_1, \ldots, Y_{n_2}$ the random variables corresponding to the defender's dice. The probability that the attacker wins is now 
%\begin{equation}
%p = \sum_{j=2}^6 P(X_{(n)} = j \land Y_{(n)} < j) = \sum_{j=2}^6 P(X_{(n)} = j) P(Y_{(n)} < j).
%\end{equation}
%If the defender has more than two troops, then the next step is to compare the $(n-1)$th order statistics %$X_{(n-1)}$ and $Y_{(n-1)}$. 

%\section{Attacking from several territories}
\subsection{Attacking from several territories}
Since one can only attack from one territory with at most three troops, it is often practical to target the same territory from several different locations. Suppose the defender has $n_d$ troops. Now when calculating the probability that the attacker occupies the territory, it must be taken into account that the first territory attacking does not win the combat but still kills $d_1$ troops. Now the second territory attacking needs to defeat $n_d- d_1$ troops, the third $n_d - d_1 - d_2$, and so on. 

Suppose the attacker is attacking from two territories with $a_1, a_2$ troops, where $a_1$ troops are attacking first. The number of defending troops is denoted with $n_d$ and the number of troops killed by the first attacking troops with $d_1$. We shall write $P(a, d, l)$ for the probability that $a$ attacking troops kill $d - l$ troops when there are $d$ troops defending, i.e. after the attack there are $l$ troops left. If we write $P(a, 0, 0) = 1$, the total probability to conquer the territory is
\begin{equation}
\label{eq:two}p = \sum_{d_1 = 0}^{n_d} P(a_1, n_d, n_d - d_1) P(a_2, n_d - d_1, 0).
\end{equation}

In the case of attack from three territories, we can use similar notation. Suppose the amounts of attacking troops are $a_1, a_2, a_3$. The probability that the attacker wins is
\begin{eqnarray}
\label{eq:three}p = \sum_{d_1 = 0}^{n_d} \sum_{d_2 = 0}^{n_d - d_1} & [P(a_1, n_d, n_d - d_1) P(a_2, n_d - d_1, n_d - d_1 - d_2) \nonumber \\ 
&  P(a_3, n_d - d_1 - d_2, 0)].
\end{eqnarray}

%\section{\label{sec:expectation}Expectation}
\subsection{\label{sec:expectation}Expectation}
Mathematically the expectation\cite{hogg:intro} (or expected value) of a function $u(X)$ of a random variable $X$ is defined as 
\begin{equation}
E[u(X)] = \sum_i u(x_i) p_i,
\end{equation}
where $p_i$ is the probability of $x_i$. 

The mean value and the standard deviation of the distribution can be defined through the expectation. 
\begin{eqnarray}
& \mu & = E[X], \\
& \sigma^2 & = E[(X - \mu)^2] = E[X^2] - \mu^2,
\end{eqnarray}
where $\mu$ is the mean, $\sigma^2$ is the variance and $\sigma = \sqrt{\sigma^2}$ is the standard deviation. The standard deviation can be interpreted as a measure of the dispersion of the points relative to $\mu$.

There are several interesting expectations in the Risk type combat. One of the most useful expectation values is the ``price'' of an attack, i.e. how many troops is likely to be lost in the combat. Again those combat sequences where exactly $e$ troops are lost can be picked from the constructed tree of all possible combat sequences. The expectation can be also calculated in the case of attack from several territories.

Suppose attacker is using $n_a$ troops. Let $p(e)$ be the probability of a successful attack where the attacker loses $e$ troops. Now the expected value of lost troops can be written as
\begin{equation}
\label{eq:attexp}E = (1 - \sum_{e=0}^{n_a-1}p(e)) n_a + \sum_{e=1}^{n_a-1} p(e)e.
\end{equation}
Note that the first sum $\sum_{e=0}^{n_a-1}p(e)$ is the probability of a successful attack, and the first term of the expectation corresponds to the situation where the attacker loses all his/her troops. 

In addition, one might be interested in the expectation of lost units when defending a territory. With that number it is possible to derive the smallest number of units for which it is likely to successfully counter an attack. For a given number of attacking units, possibly from several territories, it is possible to calculate the probability distribution of lost defending units. Suppose the number of defending units is denoted with $n_d$, and the $X_L$ is the random variable corresponding to the number of lost units. Now the expectation of the units surviving an attack is 
\begin{equation}
\label{eq:expect}E[n_d - X_L] = n_d - E[X_L].
\end{equation} In this particular case we are interested the situation where Eq. (\ref{eq:expect}) is at least one, i.e. what is the sufficient number of troops such that there is at least one troop left after the combat.

\section{\label{sec:results}Results}
\subsection{Attack from one territory}

\begin{figure}
\includegraphics[width=100mm]{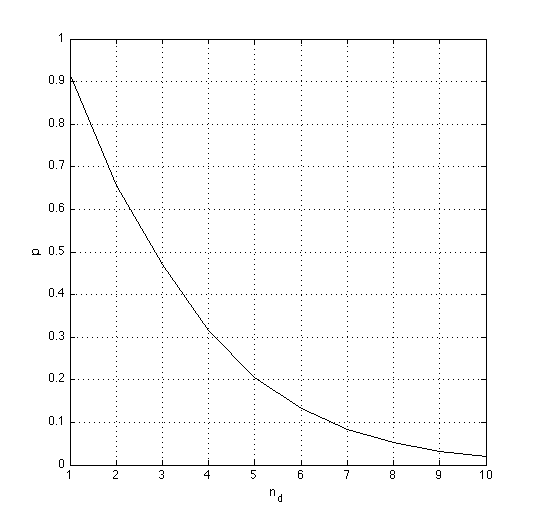}
\caption{\label{fig:single} Probability of conquering a territory with three troops when there are $n_d$ troops defending.}
\end{figure}

\begin{figure}
\includegraphics[width=100mm]{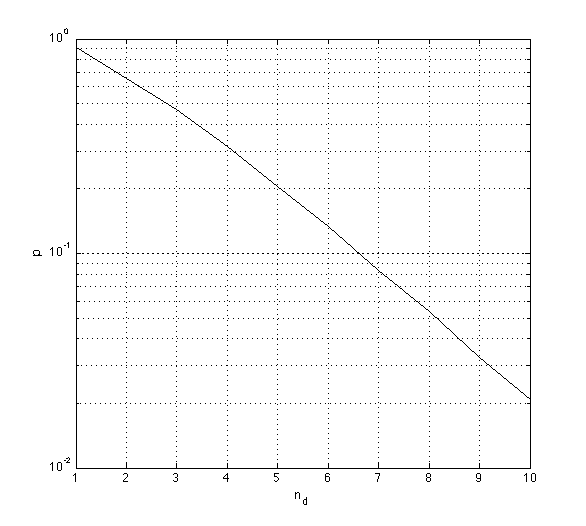}
\caption{\label{fig:singlelog} Logarithmic probability of conquering a territory with three troops when there are $n_d$ troops defending.}
\end{figure}

In the case of one-on-one attack it is quite easy to see that the probability that the attacker wins is 
\begin{equation}
p = \sum_{k=1}^5 \frac{1}{6} \frac{k}{6} = \frac{15}{36} \approx 0.42.
\end{equation}
If the attacker has two troops the probability to beat one defending troop the probability is already as high as 75\%. However, this is not a very common scene.

Suppose the attacker enters the combat with the maximum of three troops. With the methods proposed earlier in this paper one can compute the probabilities when the defender has $n_d$ troops. The probabilities can be seen in Fig. \ref{fig:single} and \ref{fig:singlelog}. When the probabilities are plotted on a logarithmic y-axis, we notice that the graph is almost a straight line. This is due to the form of Eq. (\ref{eq:ekap}) - (\ref{eq:p5}), since the total probability is a combination of products and sums of these equations. Since all possible combat sequences has to be taken into account, the resulting formula will be relatively long even in the simplest cases. Nevertheless, these sequences are easy to enumerate with a computer and the total probability can be calculated.

\subsection{Attack from several territories}

\begin{figure}
\includegraphics[width=100mm]{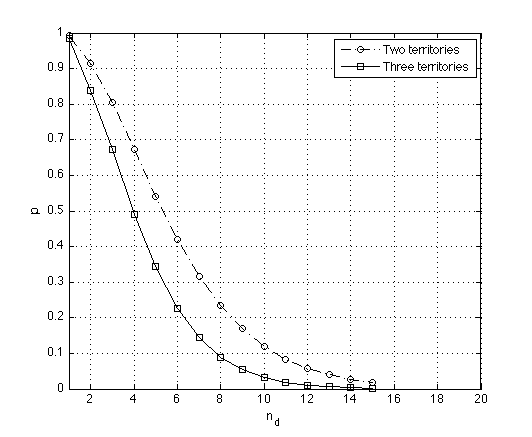}
\caption{\label{fig:sixatt} Probabilities of conquering a territory with six troops when there are $n_d$ troops defending.}
\end{figure}

In Fig. \ref{fig:sixatt} two different cases with the same amount of troops are plotted. The solid line with squares presents the probabilities of an attack from three territories with two troops in each. The dash-dot line with circles presents the probabilities of an attack from two territories with three troops in both. From the figure one can see the obvious result that it is better to attack with more troops on one round than to split the same amount into a greater number of territories.

\subsection{Expectation of lost troops in an attack}

\begin{figure}
\includegraphics[width=100mm]{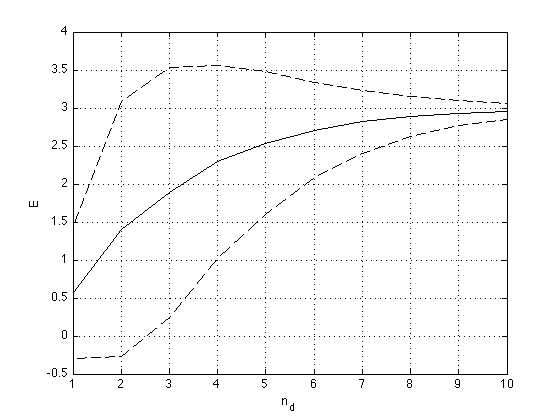}
\caption{\label{fig:attlose} Expectation of lost troops when attacking from a single territory with three troops. The dashed lines denote the boundaries given by one standard deviation.}
\end{figure}

The expectation $E$ as introduced in Eq. (\ref{eq:attexp}) when attacking from a single territory with three troops is drawn in Fig. \ref{fig:attlose} along with the one $\sigma$ boundaries. The expectation of lost troops approaches the limit three when the number of defenders $n_d$ increases. In this particular case the standard deviation $\sigma$ reaches it's maximum when there are three troops defending.

\subsection{Expectation of surviving defending troops}

\begin{figure}
\includegraphics[width=100mm]{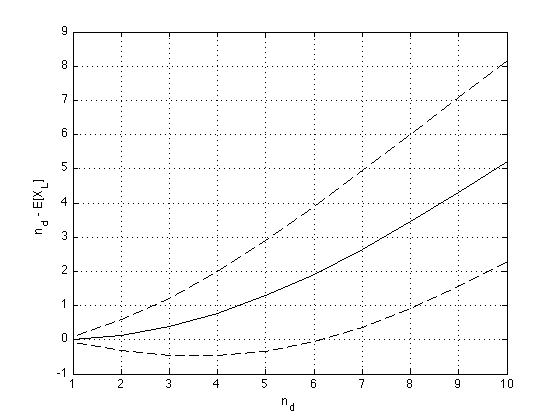}
\caption{\label{fig:defleft} Expected number of surviving defending troops when 6 troops are attacking from two territories. The dashed lines denote the boundaries given by one standard deviation.}
\end{figure}

The smallest value for $n_d$ for which Eq. (\ref{eq:expect}) gives at least one, are 3, 5, and 7, when the attacker is using 3, 6, and 9 troops, respectively. However, when looking the probabilities, in the two latter cases $n_d$ should be increased by one to reach 0.5 for the probability of countering the attack. 

%Since the distribution of the surviving units has finitely many nonzero entries and can be fully calculated, it is also possible to find the standard deviation\cite{hogg:intro}. 
An attack of six troops to a territory defended by $1 \ldots 10$ troops is calculated. The expectation of the surviving troops and boundaries given by one standard deviation is drawn in Fig. \ref{fig:defleft}. 
%As can be seen the standard deviation is relatively large when compared to the expected value.

%\begin{figure}
%\includegraphics[width=90mm]{defstd6.png}
%\caption{\label{fig:defleftstd} Standard deviation of the surviving defending troops under an attack of six troops.}
%\end{figure}

\section{Conclusions}

Probabilities to conquer a territory were calculated in several different situations, and some lower bounds for a sufficient number of defending units were derived using the expected value of surviving units. However, the probability to lose a territory to six troops with ten defending troops is still as high as 12\%. With this kind of probability distributions, it can be said that the game is almost solely based on dice rolling as far as the players avoid the coarsest strategic flaws. Moreover, the same result is suggested by the relatively large standard deviations in the distributions.

%\appendix
%\section{}

%\begin{acknowledgments}
%\end{acknowledgments}

\bibliographystyle{amsplain}
\bibliography{risk}

\providecommand{\bysame}{\leavevmode\hbox to3em{\hrulefill}\thinspace}
\providecommand{\MR}{\relax\ifhmode\unskip\space\fi MR }
\providecommand{\MRhref}[2]{%
  \href{http://www.ams.org/mathscinet-getitem?mr=#1}{#2}
}
\providecommand{\href}[2]{#2}
\begin{thebibliography}{1}

\bibitem{daviau:risknew}
Rob Daviau, \emph{Risk rulebook (revised edition)}, Hasbro, 2008.

\bibitem{hogg:intro}
R.V. Hogg and A.T. Craig, \emph{Introduction to mathematical statistics},
  Macmillan, 1970.

\bibitem{lamorisse:risk}
Albert Lamorisse, \emph{Risk rulebook}, Parker Brothers, 1957.

\end{thebibliography}

\end{document}